\documentclass[twocolumn,showpacs,preprintnumbers,amsmath,amssymb]{revtex4}

\usepackage{graphicx}
\usepackage{dcolumn}
\usepackage{bm}

\begin{document}

\preprint{APS/123-QED}

\title{Vortex Turbulence in Linear Schroedinger Wave Mechanics}

\author{Tzihong Chiueh}
\email{chiuehth@phys.ntu.edu.tw}
\author{Tak-Pong Woo}
\email{bonwood@scu.edu.tw}

\author{Hung-Yu Jian}
\email{hyj@phys.ntu.edu.tw}

\author{Hsi-Yu Schive}
\email{b88202011@ntu.edu.tw}
\affiliation{%
1. Department of Physics,\\
National Taiwan University, 106, Taipei,
Taiwan\\ 2. Center
for Theoretical Sciences, National Taiwan University, 106, Taipei,
Taiwan\\ 3. Center
for Quantum Sciences and Engineering, National Taiwan University,
106, Taipei, Taiwan\\4. LeCosPa, National Taiwan
University, 106, Taipei, Taiwan\\5. Department of Physics, Soochow University
}


\date{\today}

\begin{abstract}
Quantum turbulence that exhibits vortex creation, annihilation and interactions
is demonstrated as an exact solution of the time-dependent, free-particle
Schroedinger equation evolved from a smooth random-phased initial
condition. Relaxed quantum turbulence in 2D and
3D exhibits universal scaling in the steady-state energy spectrum
as $k^{-1}$ in small scales.  Due to the lack to dissipation, no evidence of Kolmogorov-type
forward energy cascade in 3D or inverse energy cascade in 2D is found, but
the rotational and potential flow components do approach equi-partition in the scaling regime.
In addition, the 3D vortex line-line correlation exhibits universal behavior,
scaled as $\Delta r^{-2}$, where $\Delta r$ is the
separation between any two vortex line elements, in fully developed turbulence.
We also show that quantum vortex is not frozen to the
matter, nor is the vortex motion induced by other vortices via
Biot-Savart's law. Thus, quantum vortex is actually a nonlinear wave,
propagating at a speed very different from a classical vortex.
\end{abstract}

the Physics and Astronomy
\maketitle

\section{\label{sec:level1}Introduction}

Vortices are probably the most common features in classical
fluid turbulence, regardless of compressible or incompressible flows.
Likewise, quantized vortex lines and tangles not only are frequently observed
but also play vital roles in quantum turbulence [1,2,3]. Quantized vortices
also exist in type II superconductors, representing defects in
superconductivity within which resistivity is finite [4]. Recent
interests in quantum vortices have been motivated by the intriguing dynamics in
Bose-Einstein Condensation [5.6,7].  Particularly interesting is the
existence of Kelvin waves on vortex lines [8,9]. Kelvin waves are helical
waves and believed to be responsible for coupling quantum turbulence to dissipation.

While two dimensional quantum vortices can be created or annihilated in pair,
three dimensional vortex lines can change their topology through reconnection.
These intriguing dynamical behaviors have previously been reported with a simple model [10].
However, it is not clear how quantum turbulence may arise from abundant quantum
vortex excitations, with the dynamics of each vortex obeying what is described by that work.

Turbulence can occur when an abundance of free energy is available.
External forcing is one option to provide the free energy, yielding
the driven quantum turbulence. A configuration unstable to large scale fluctuations,
such as that due to gravity [11], is an alternative to make the free energy
available, creating the relaxed turbulence.  Driven turbulence requires an energy sink or
dissipation so as to achieve detailed balance in a steady state.
Relaxed turbulence on the other hand does not necessarily
involve in dissipation.  But when it does, turbulence will only be a transient phenomenon
and will eventually decay away.  On the other hand, dissipationless relaxed turbulence that obeys Hamiltonian
dynamics with
constant energy is a rather uncommon phenomenon in a macroscopic world, since when all degrees of freedom are excited the smallest scale in real systems can often couple to dissipation.  Moreover in
most theoretical models of turbulence, integration of dynamical equations must often involves numerical dissipation so as to warrant numerical stability, thereby rendering investigation of dissipationless turbulence difficult.

Dissipationless dynamics can nevertheless exist in isolated microscopic systems.
How can a microscopic dynamical system evolve to a turbulence state?
A turbulent system is normally conceived to be chaotic in detail but universal (insensitive to initial conditions) on average. It is unquestionable that a many-body system with nonlinear interactions can be chaotic. However, it is unclear how a universal state can dynamically arise from a Hamiltonian system.  One of the goals of this work is to
demonstrate such a universal non-equilibrium state in a Hamiltonian system.

Another important feature which may also exist in turbulence is the
coherent structure [12]. Coherent structures are comparatively
longer-lived than other more common random flows in turbulence,
and coherent structures may even possess correlation among different
length scales of great disparity. The latter property suggests
that local scale coupling, which is the basis of the Kolmogorov turbulence
cascade scenario, does not necessarily hold, and can lead to intermittency and
deviation from the universal turbulence spectrum derived from the
mean-field theory [13]. Quantum vortices may be regarded as such coherent
objects existing in quantum turbulence, and we will show that their existence can alter
the turbulence small-scale energy spectrum.

Inspired by the concept developed for classical fluid turbulence,
we are motivated to examine quantum turbulence in the framework
of vortex creation and annihilation. Due to the strong nonlinearity in
turbulence we shall in principle investigate this problem numerically.
However, numerical simulations can often yield uncontrollable
numerical dissipation that depends on the adopted numerical
schemes. Because the quantum dynamics obeys certain exact
symmetries, such as the unitary condition,
which can be difficult to be strictly reinforced in numerical simulations, we
therefore study a simplest, soluble but non-trivial case in this
work --- a system described by the Schroedinger equation of free
particle with random-phased initial conditions.  In the context of superfluid,
our system corresponds to the regime where the vortex healing length diverges.

This system in
wavefunction formulation is linear and exactly soluble.
However, in the fluid formulation, this linear quantum system can
be strongly nonlinear. When compared with the nonlinear
Schroedinger equation system, the difference is simply the absence of
polytropic fluid pressure in the linear Schroedinger case. The
polytropic fluid pressure is responsible for sound emission outside
vortex cores and important only on large scale.  Due to the existence of
the polytropic fluid pressure, in addition to
the forward cascade of kinetic energy the occupation number per mode
can also exhibit an inverse cascade in favor of large-scale excitation.
(In the linear Schroedinger case, the occupation number per mode is conserved
and the inverse cascade is prohibited.)
But important features of quantum turbulence, such as vortices, are of
small scale. It is therefore anticipated that the linear
Schroedinger equation should be able to capture these small-scale features.

This reasoning is strengthened by the
following observation of duality for the nonlinear Schroedinger system.
At the very location of vortex where
the fluid formulation becomes strongly nonlinear and the density
vanishes, the system becomes linear in the wave formulation. The same
argument also applies to Kelvin waves which are located at the
low-density vortex cores.
By contrast, the strongly nonlinear regime of a nonlinear Schroedinger system
in the wave formulation can be approximated in the fluid formulation by
a linear equation for sound waves. The duality
conjugation of strong coupling regimes between these two different
formulations leads us to believe that our linear wave model is
essential to understand quantum vortices and has generic
relevance to quantum turbulence of the full problem described by
a nonlinear Schroedinger equation.

This paper is organized as follows. Sec.(II) presents the fluid
formulation, an alternative to the conventional wave
formulation, and the connections of the two in terms of
vortices. We give the numerical results of quantum turbulence
prior to full excitation of vortices in Sec.(III). Vortices in fully
developed turbulence are addressed in Sec.(IV). Our conclusion
is given in Sec.(V). Throughout this work, we let $\hbar=1$
and mass $m=1$.

\section{Fluid Formulation}

To show the nonlinearity in the fluid variables of the
free-particle Schroedinger equation, we cast the wave equation
into a set of fluid equations. Let the complex wavefunction be
represented by two real functions, $\Psi=f e^{iS}$, with $f$
representing the real amplitude and $S$ the real phase. We
further let the fluid density be $\rho\equiv f^2$ and the fluid
velocity ${\bf v}\equiv Im(\Psi^*\nabla\Psi)/\rho=\nabla S$. The
Schroedinger equation then becomes
\begin{equation}
{\partial f^2\over\partial t}+\nabla\cdot(f^2\nabla S)=0
\end{equation}
and
\begin{equation}
{\partial\nabla S\over\partial t}+\nabla[{(\nabla S)^2\over
2}-{\nabla^2 f\over 2f}]=0.
\end{equation}
The first equation represents the conservation of mass. We also
replace the original $\nabla(\partial S/\partial t)$ by
$\partial(\nabla S)/\partial t$ in the second equation, representing the Euler
equation for fluid momentum.

On the other hand, when $S$ becomes a multi-valued function, vortices appear and rotational flows set in. The rotational flow is induced by vortices via Biot-Savart law. The multi-valued $S$ can appear when $\rho$ locally vanishes and the quantum potential $\nabla^2 f/2f$ diverges. Therefore vortices are always located
at regions where the wavefunction vanishes [10,14]. Since the
wavefunction is complex, this condition demands both real part $R$
and imaginary part $I$ to vanish simultaneously, i.e., $\Psi =
({\bf x}-{\bf x}_0)\cdot\nabla R +i ({\bf x}-{\bf x}_0)\cdot\nabla
I$. Normally $\nabla R$ and $\nabla I$ are not aligned at ${\bf
x}={\bf x}_0$, and the intersection of the two non-parallel
surfaces $R=0$ and $I=0$ forms a line in 3D, the vortex line. In 2D,
the intersection is a point vortex.

The local density near the vortex can be expressed as $\rho=
[({\bf x}-{\bf x}_0)\cdot\nabla R]^2 +[({\bf x}-{\bf x}_0)\cdot\nabla
I]^2$, which vanishes quadratically. Upon diagonalization it becomes
$\rho= a^2(x'-x'_0(t))^2+b^2(y'-y'_0(t))^2$ where $a$ and $b$ are constants. Without loss of generality, we can therefore let $\Psi\sim a(x-x_0(t))+ib(y-y_0(t))$
in an instantaneous diagonal frame where the primes in the above expression is dropped for convenience. From this local expression of wavefunction, we find that
\begin{equation}
\Psi=\sqrt{a^2(x-x_0)^2+b^2(y-y_0)^2}e^{i\phi}
\end{equation}
where $S=\phi$ which is the space rotation angle around the vortex
line. Thus, isolated vortex is generally quantized with a unit angular
momentum. (Vortex quantized with an integer multiple of unit
angular momentum is also allowed, but it is much less common to be
excited in dynamical evolution.) Near the vortex line the
leading-order velocity
\begin{equation}
{\bf v}=\nabla S={ab r\over \rho}\hat\phi,
\end{equation}
where ${\bf r}={\bf x}-{\bf x}_0$. The velocity diverges at the vortex and
the singular vorticity can be defined as
\begin{equation}
{\bf\omega}\equiv\nabla\times\nabla S\equiv{\hat z\over \pi r_c^2}
\lim_{c\to 0}\int_c \nabla S\cdot d{\bf l} =\delta(r)\hat z,
\end{equation}
where the subscript $c$ represents a circular contour along $\phi$
and $r_c$ is the radius of the circle $c$.
This definition of curl operation makes the singular vorticity circular symmetric. In fact when the density is not circularly symmetric, the leading order velocity also contains compression $\nabla^2 S$, which also diverges,
\begin{equation}
\nabla^2 S = {ab(b^2-a^2)\sin(2\phi)\over r^2(a^2\cos^2(\phi)
+b^2\sin^2(\phi))^{2}}.
\end{equation}
One can further derive that
\begin{equation}
S=\tan^{-1}({b\over a}\tan(\phi)),
\end{equation}
to the leading order near the vortex, and we can uniquely separate
the rotational component $S^r$ from the potential component $S^p$,
with $S^r=\phi$ and $S^p= tan^{-1}[(b/a)\tan(\phi)]-\phi$ at the vortex.

In addition, vortex can move at a finite velocity perpendicular
to the vortex line. Since the vortex motion is locally
two-dimensional, we shall consider the 2D case in the following
local analysis. The vortex motion can be determined from the
density evolution equation, since the vortex is tied to
$\rho=0$. Substituting the local expression of ${\bf v}$ and
$\rho$ into Eq.(4), we find that to the leading order
$\nabla\cdot(\rho{\bf v})=ab\nabla\cdot (r\hat\phi)=0$. Let
$\rho=\rho_0+\delta\rho$ and ${\bf v}={\bf v}_0+\delta{\bf v}$,
where $\rho_0$ and ${\bf v}_0$ are the leading order density and
velocity given above. We then obtain
\begin{equation}
\nabla\cdot(\rho_0\delta{\bf v}+\delta\rho{\bf v}_0) = ({\bf
r}\cdot\nabla R)\nabla^2 I - ({\bf r}\cdot\nabla I)\nabla^2 R
\end{equation}
to the next order. Since this equation is a linear equation in $x$
and $y$, we can regard it as the convection of the density null with a
finite velocity ${\bf w}$; that is, ${\bf w}\cdot\nabla\rho_0$, where
${\bf w}=d{\bf x}_0(t)/dt$, describing the instantaneous
velocity of a vortex. A straightforward algebra shows that
\begin{equation}
{\bf w}={{Re[\nabla^2\Psi\nabla\Psi^*]\times\hat z}\over{2
\nabla\Psi\times\nabla\Psi^*\cdot\hat z}},
\end{equation}
where the gradient $\nabla$ is two-dimensional perpendicular to the local vortex line. Note that the propagation velocity is primarily determined by the Laplacian
of wave function. This expression is the same
as that obtained in [10] with a somewhat different approach.

With this expression for ${\bf w}$, we may inquire whether ${\bf
w}$ is the same as, or can be reduced from, the material velocity $\delta{\bf v}$. This question is important since if it does the vortex will be frozen
to the quantum fluid and gets advected as the matter does, much like the classical
vortex in an isentropic classical fluid. If it
does not, the quantum vortex is then a wave pattern, propagating with
its own velocity. As far as we are aware of, this important issue
has not been addressed in the literature.

Since the vorticity $\omega(\equiv\nabla\times{\bf v})$ is a Dirac
$\delta$-function, which has circular symmetry, the zeroth-order
velocity, being along the circular direction, can not displace the vortex.
That is, ${\bf v}_0\cdot\nabla\omega=0$. It is the next-order velocity
$\delta{\bf v}$ that can provide a non-vanishing contribution to vorticity
advection. Since the leading-order velocity is singular, the next-order velocity
can only be obtained by regularization.
Regularization of velocity can be performed by averaging the continuity equation, Eq.(1), over an infinitesimally small disk around the vortex:
\begin{equation}
{\partial\langle\rho\rangle\over\partial t}+\nabla\cdot\langle\rho\nabla S\rangle=0.
\end{equation}
The continuity equation is used to defined the advection velocity because it ensures that the averaged material the vortex is advected by an effective material velocity. Such an average is equivalent to a local angular average, and the regularized velocity $\overline{\delta{\bf v}}$ can be shown to be
\begin{equation}
\overline{\delta{\bf v}}\equiv{{\langle\delta(\rho\nabla S)\rangle}
\over{\langle\rho\rangle}}
= {{Im[\nabla\Psi^*\cdot(\nabla\nabla-{\bf I}\nabla^2/2)\Psi]}\over
{2|\nabla\Psi|^2}},
\end{equation}
after a straightforward algebra. Here the identity matrix ${\bf I}$ is two dimensional and so is the gradient $\nabla$.
The material velocity $\overline{\delta{\bf v}}$ near the vortex is proportional to the quadrupole expansion of the wave function and apparently different from the vortex velocity ${\bf w}$ which is proportional to the Laplacian of the wave function. It evidences that the vortex is not frozen to the matter and must therefore be a wave pattern.

Finally, we may want to examine whether the vortex motion arises from
the induction velocity generated by other vortices via the
Biot-Sarvart law.  Note that the induction velocity is incompressible, whereas
the material velocity described by Eq.(11) is composed of both compressible and
incompressible components and therefore the induction velocity is different from
the material velocity.  This question is important as it has often
been assumed so in the literature.  We find that the answer is again negative,
as can be illustrated from the following simple case.

Consider a two-dimensional free-particle wave function in cylindrical coordinate $(R, \phi)$,
$\Psi=c_0 - J_1(kR) e^{i(\phi-k^2 t/2)}$, where $c_0$ is
a positive constant, $k^2/2$ the energy eigenvalue and $J_1$ the first-order Bessel function.
The density $|\Psi|^2$ is stationary in a counter-clockwise
rotating frame with an angular velocity $Rk^2/2$ around $R=0$.
This system contains only two vortices of opposite signs
near the first peak of $J_1$ when $J_1(kR_1) > c_0$ and $J_1(kR_2)
< c_0$, where $R_1$ and $R_2$ are the first and second peaks of
$J_1$. Since the two density nulls are also stationary in the same rotating frame,
the two vortices travel together counter-clockwise with the same angular velocity.
If the vortex motion were to be due to the induction of the other vortex, the two
should have traveled with a speed depending on their separation.
One may change the vortex separation by adjusting $c_0$. However the angular velocity
of the vortex pair remains the same under such a change, meaning
that the vortex motion is not caused by the mutual induction.

The nature of a quantum vortex is therefore very different from a classical
vortex of an isentropic classical fluid, for which the fluid dynamics is described
by the same mass continuity equation as Eq.(1) and a similar momentum equation as Eq.(2). The classical vortex is frozen to the matter. However, the quantum vortex is a nonlinear wave, whose propagation is controlled by the local Laplacian of wave function. As a consequence, the dynamics of quantum
vortex, or that of the rotational velocity component, cannot be
described by the fluid formulation given by Eqs.(1) and (2) alone for a potential flow. When vortices appear in a quantum fluid, an additional equation of motion for vortex is needed in the fluid formulation to evolve the whole system.

To trace the origin of this issue, let us return to the momentum equation of
the fluid formulation, Eq.(2). We shall prove from the vortex motion
that the time derivative and the curl operation acting on the rotational (incompressible) components
of velocity, $\nabla S^r(={\bf v}^r)$, do not commute.  (Note that the velocity field also contains a
potential (compressible) component, $\nabla S^p$, which is regular and not associated with vortices.)
First, we shall show that
$\nabla$ and $(\partial/\partial t)$ commute when acting on $S^r(\equiv tan^{-1}[(y-y_0(t))/(x-x_0(t))])$. Let the instantaneous
velocity vector ${\bf w}(=d{\bf x}_0/dt)$ be chosen in the x-direction, ${\bf r}_0=0$ and $\phi$ be the angle around $r=0$. The equality
\begin{equation}
\nabla(\partial S^r/\partial t)
=(w/r^2)[\cos(2\phi)\hat y-\sin(2\phi)\hat x]=\partial(\nabla S^r)/\partial t
\end{equation}
holds near the vortex, ensuring that $\nabla$ and $\partial/\partial t$ commute. Hence Eq.(2) is a valid equation derived from the Schroedinger equation even in the presence of rotational component.
Second, the quantity $\partial S^r/\partial t =-\sin(\phi)/r$
is a single-valued function, and therefore $\nabla(\partial S^r/\partial t)$
is curl-free. So will $\partial\nabla S^r/\partial t$ be curl-free.  Upon taking a curl
operation on Eq.(2), both sides are identically zero.

On the other hand, we have the vortex equation of motion
\begin{equation}
{\partial{\nabla\times\nabla S^r}\over\partial t}=-{\bf w}(t)\cdot\nabla(\nabla\times\nabla S^r),
\end{equation}
where the vorticity $\nabla\times\nabla S^r=
\delta({\bf r}-{\bf r}_0(t))$. Combining with the fact that
$\nabla\times(\partial\nabla S^r/\partial t)=0$, we have
the commutation relation
\begin{equation}
[\nabla\times,\partial/\partial t]\nabla S^r={\bf w}\cdot\nabla(\nabla\times\nabla S^r),
\end{equation}
which is non-vanishing. This completes our proof.

From the viewpoint of dynamical evolution, we find that the fluid formulation of Quantum Mechanics,
Eqs.(1) and (2), are inadequate when vortices exist in the system.
A third equation must be supplemented, with the
above commutation relation, Eq.(14), or the vortex equation, Eq.(13). This new equation updates the
positions of vortices, and the instantaneous rotational velocity $\nabla S^r$ outside the vortex can be determined by the Biot-Savart law of
induction. Since the vortex velocity ${\bf w}$, Eq.(9), depends also on
the potential component of the flow $\nabla S^p$, the rotational component
$\nabla S^r$ is thereby coupled to the potential component.
On the other hand, the potential component $\nabla S^p$ is coupled to the
rotational component through the momentum equation, Eq.(2), since $\nabla S$
in this equation contains both $\nabla S^r$ and $\nabla S^p$. It thus
follows that these two topologically different components of flow are non-trivially coupled.

\section{Time-Dependent Solutions of Quantum Turbulence}

In order to understand the development of vortex creation and the subsequent evolution in quantum turbulence, we solve the time-dependent solutions of the free-particle Schrodineger equation in a periodic box of $2048^2$ grids for 2D and of $1024^3$ grids for 3D. The solution can be given analytically through the Fourier-mode propagator, and the periodic computational box used here serves only to illustrate the turbulence solutions.

\subsection{Initial Conditions and Potential Flow Turbulence}

We define the potential (compressional) flow in the Fourier
space as ${\bf v_k}^p\equiv \hat k\hat k\cdot {\bf v_k}$. The
rotational (incompressible) flow is then defined as
${\bf v}^r\equiv{\bf v}-{\bf v}^p$.
We choose the initial condition such that $\rho=1$ throughout
the entire domain and let the phase be randomly distributed only
on the large scale. Specifically, the Fourier components of $S$
obey a Gaussian random distribution centering at $|{\bf k}|=0$
with the variance $|\Delta{\bf k}|$ equal to $20\times 2\pi/L$ for
2D and $10\times 2\pi/L$ for 3D, where $L$ is the domain
size. Thus the initial flow energy is confined on large scales.

We construct the time-dependent solution of the linear Schroedinger
equation by using the propagator in Fourier space. Each Fourier
component of mode ${\bf k}$ is propagated forward in time by
multiplying the phase factor $e^{it{\bf k}^2/2}$ to the initial
Fourier mode. The updated wavefunction is then used to construct
the fluid variables, $\rho$ and ${\bf v}$, for exploration of
quantum fluid turbulence.

Due to the nonlinearity of the fluid variable, the flow energy is
able to couple from large scales to small scale. At some point in
time, the smallest scale features in the computation domain are
fully excited by the nonlinear coupling and the vortex flow
suddenly appears, as shown in the velocity spectrum (Fig.(1)). It
is noted that near the onset of the vortex flow, the small-scale
potential flow is fully excited up to the smallest available scale and the
spectrum of the potential flow develops a power law behavior,
$k^{-1}$ for 2D and $k^{-2}$ for 3D. This behavior lasts only for
a short moment before the rotational flow begins to get excited.

\begin{figure}
 \begin{center}
  \leavevmode
  \vspace{1cm}
  \includegraphics[width=8cm,angle=0]{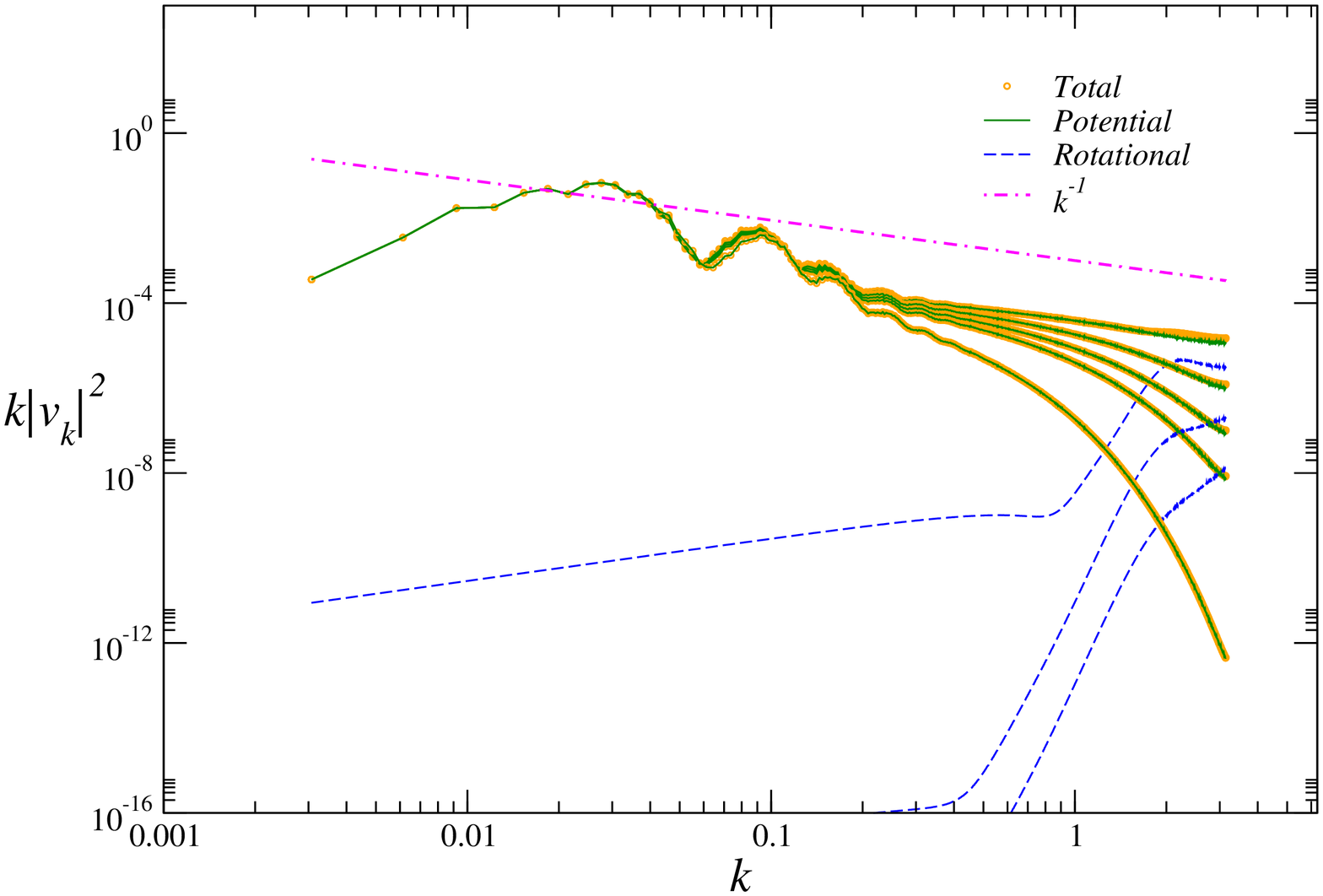}
  \includegraphics[width=8cm,angle=0]{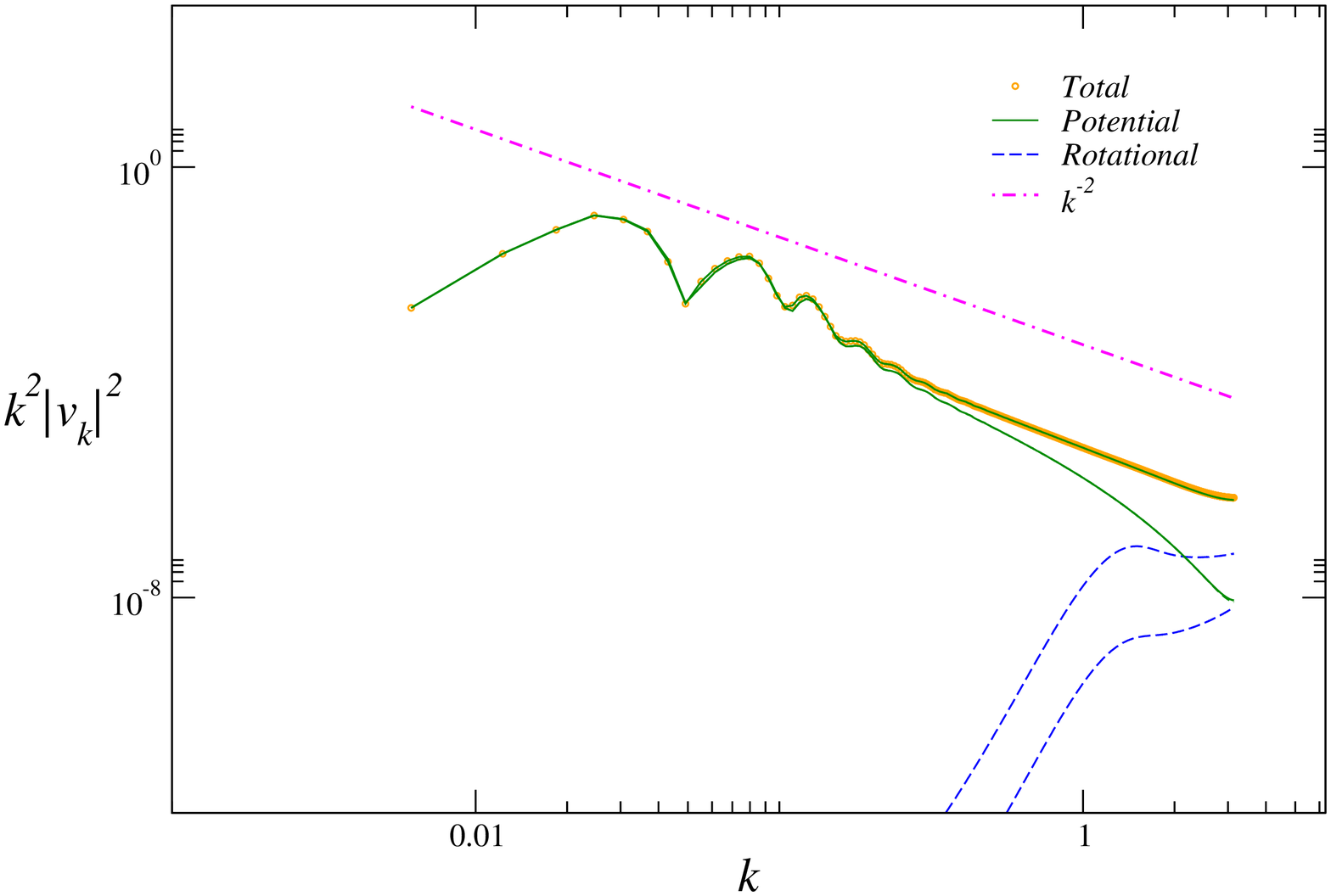}\\
  \caption{Evolution of power spectra for potential and rotation flows before
the onset of vast excitation of rotational flows for 2D (upper panel) and 3D
(lower panel).
Cascade of potential flow energy from large scales to small scales is
apparent as the small-scale
power progressively rises. In the mean time, increasing numbers of vortex pairs
are forming during this stage, signified by the small scale excitation in
the rotational flow power spectrum.}\label{fig:Velocity_Spectrum}
 \end{center}
\end{figure}

We shall pause here and offer an explanation for the power-law
spectrum of the potential flow. In a quantum system studied here,
the only dimensional constant is the Planck constant, which has
dimension $(length)(velocity)$. It then follows that the velocity
difference $|\Delta{\bf v}|$ of the potential flow over a distance
$|\Delta{\bf r}|$ scales as $|\Delta{\bf r}|^{-1}$, as a result of
uncertainty principle. The Fourier component $|\Delta {\bf
v}_k|\sim k^{-1}$ for 2D and $|\Delta {\bf
v}_k|\sim k^{-2}$ for 3D, yielding a spectral density $\sim k^{-1}$ for
2D and $k^{-2}$ for 3D. Note that the development of this
power-law behavior is through cascades, where modal energy is
transferred from large scales to small scales by local coupling in
$k$ space. This nonlinear coupling is similar to cascades in
classical fluid turbulence.

To gain a better picture of what is going on at the onset of
vortex creation, we shall turn to the real-space flow. Figure (2)
shows the local rise of 2D potential-flow velocity due to a large
local gradient in quantum potential, corresponding to the
situation where $R=0$ and $I=0$ surfaces are locally tangent to
each other. Shortly after, two streams of counter-rotating flow develop
about the two sides of the peak potential flow velocity to form
a tightly bound vortex pair, corresponding to the critical crossing
of the two
surfaces. The two counter-rotating flow streams grow in strength until
their circulations are quantized before the two vortices can
separate from each other to become isolated vortices. In 3D, vortex
creation is through formation of a vortex loop of infinitesimal radius. The
physical picture is similar to what happens for a vortex pair in
2D. Not until the rotational velocity grows to the strength where
the circulation is quantized can the vortex loop expand to a
finite radius.

\begin{figure}
 \begin{center}
  \leavevmode
  \vspace{1cm}
  \includegraphics[width=10cm,angle=0]{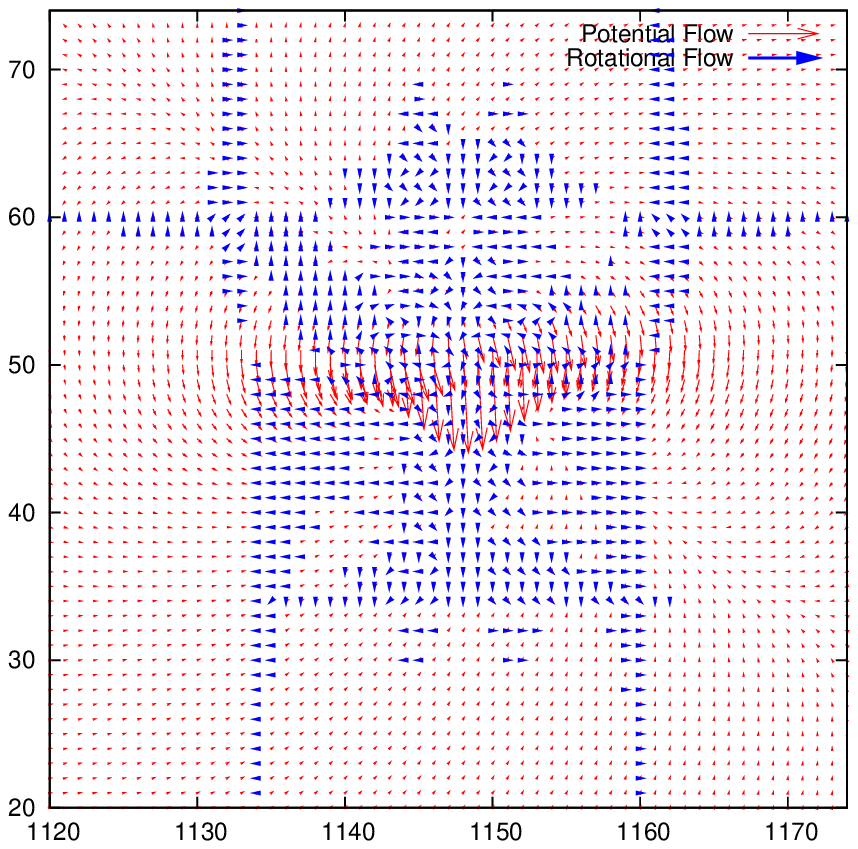}
  \includegraphics[width=10cm,angle=0]{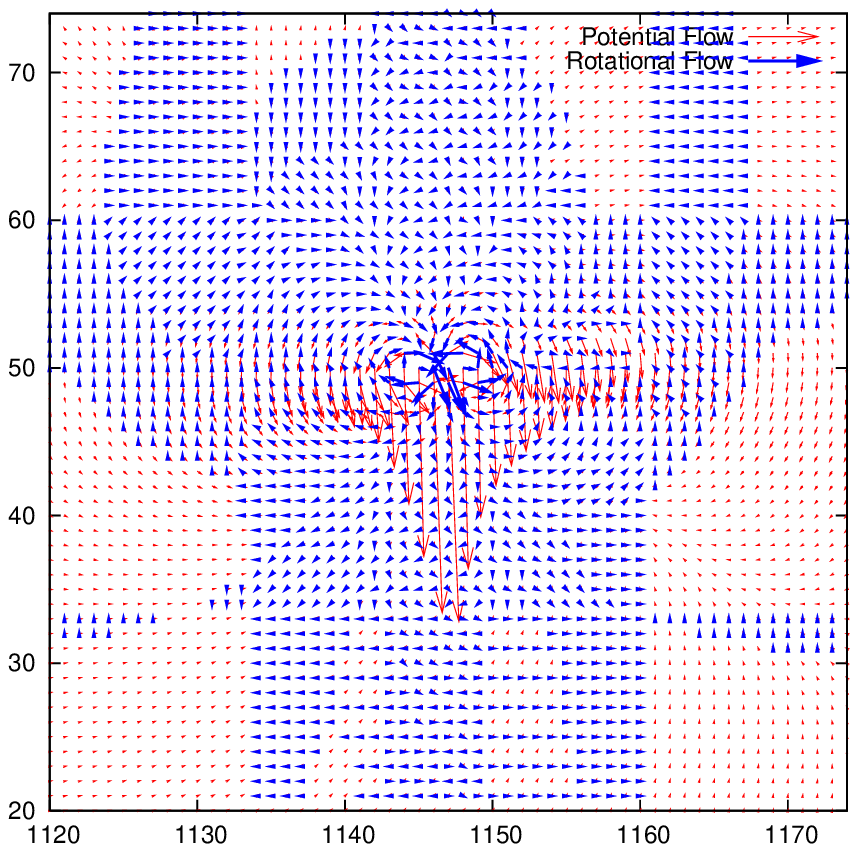}\\
  \caption{The real-space flow configuration during vortex-pair creation
in 2D.
Upper panel shows the sole existence of potential flow (red) prior to the pair
creation, where the potential flow locally develops a large velocity gradient.
 Lower panel shows the birth of a vortex pair, where counter-rotating flows
(blue) grow in strength. Not until the rotational flow velocity
grows to the magnitude where the circulation is quantized can the
pair separate as isolated vortices.}
  \label{fig:2D_vortex_merger}
 \end{center}
\end{figure}

\subsection{Vortex Turbulence}

Once created, vortices of opposite signs can move away from each
other in 2D and vortex loops can expand in 3D. Starting from our relatively smooth initial condition,
after the first vortex pair is created, many more vortex pairs can soon be generated,
tapping the free energy in the potential flow. In Figs.(3) and (4), we show
a snapshot of point vortices and vortex lines in our 2D and 3D computation domains.

\begin{figure}
 \begin{center}
  \leavevmode
  \includegraphics[width=10cm,angle=0]{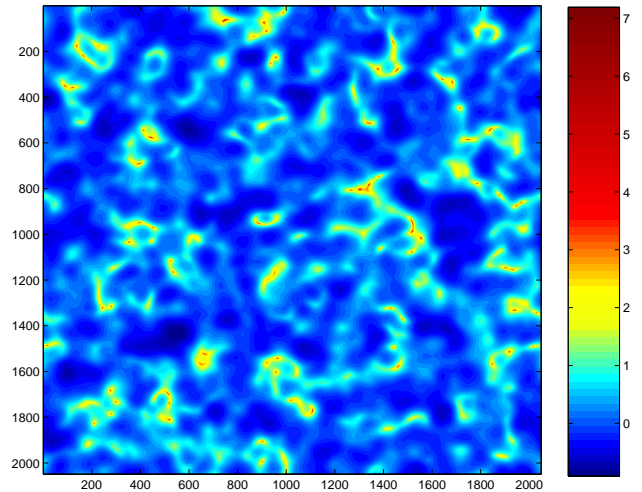}\\
  \caption{Real-space image of point vortices in steady-state 2D
turbulence. These point vortices are captured by plotting the
inverse density of extremely large value. The color code is in
logarithmic scale.}\label{fig:2048_sqr}
 \end{center}
\end{figure}

\begin{figure}
 \begin{center}
  \leavevmode
  \includegraphics[width=8cm,angle=0]{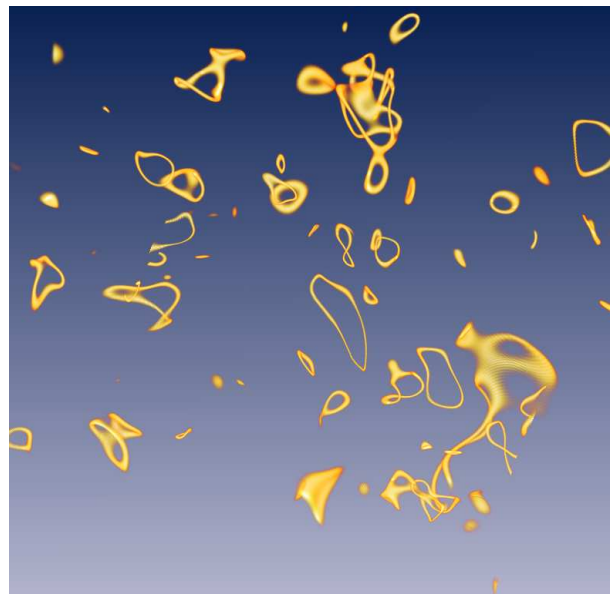}\\
  \caption{Real-space image of tangled vortex tubes in steady-state 3D
turbulence. These vortex tubes are captured by plotting the
constant-density surfaces of very low
density.}\label{fig:1024_cube}
 \end{center}
\end{figure}

The turbulence spectrum long after the onset of vortex
fluctuations is shown in Fig.(5). The flow eventually is
populated with vortex fluctuations and the spectrum approaches a
steady state that also largely obeys a power law as $k^{-1}$ in
both 2D and 3D cases. (The high-$k$ upturn in the spectrum is an artifact of
the finite-size grid and will be
addressed later.) Despite the potential flow turbulence is
suppressed in small scale after the rotational flow is vigorously
excited, the potential and rotational flow turbulence energies are
actually in equi-partition. Any vector field can be decomposed to
the potential component and rotational component as $\nabla\phi$
and $\hat z\times\nabla\psi$, respectively, in 2D, but as
$\nabla\phi$ and ${\bf A}$, constrained by $\nabla\cdot{\bf A}=0$,
in 3D. That is, a 2D flow has two degrees of freedom arising from
the independent $\phi$ and $\psi$, and a 3D flow has 3 degrees of
freedom given by one $\phi$ and and two independent components of
${\bf A}$. If these
components are non-correlated, equi-partition between potential
flow turbulence and rotational flow turbulence is naturally
expected. In 2D, the equi-partitioned ratio of potential flow
energy to rotational flow energy is $1$, and in 3D the ratio
becomes $1/2$. These behaviors are what have been observed in our
2D and 3D cases.

\begin{figure}
 \begin{center}
  \leavevmode
  \vspace{1.0cm}
  \includegraphics[width=8.cm]{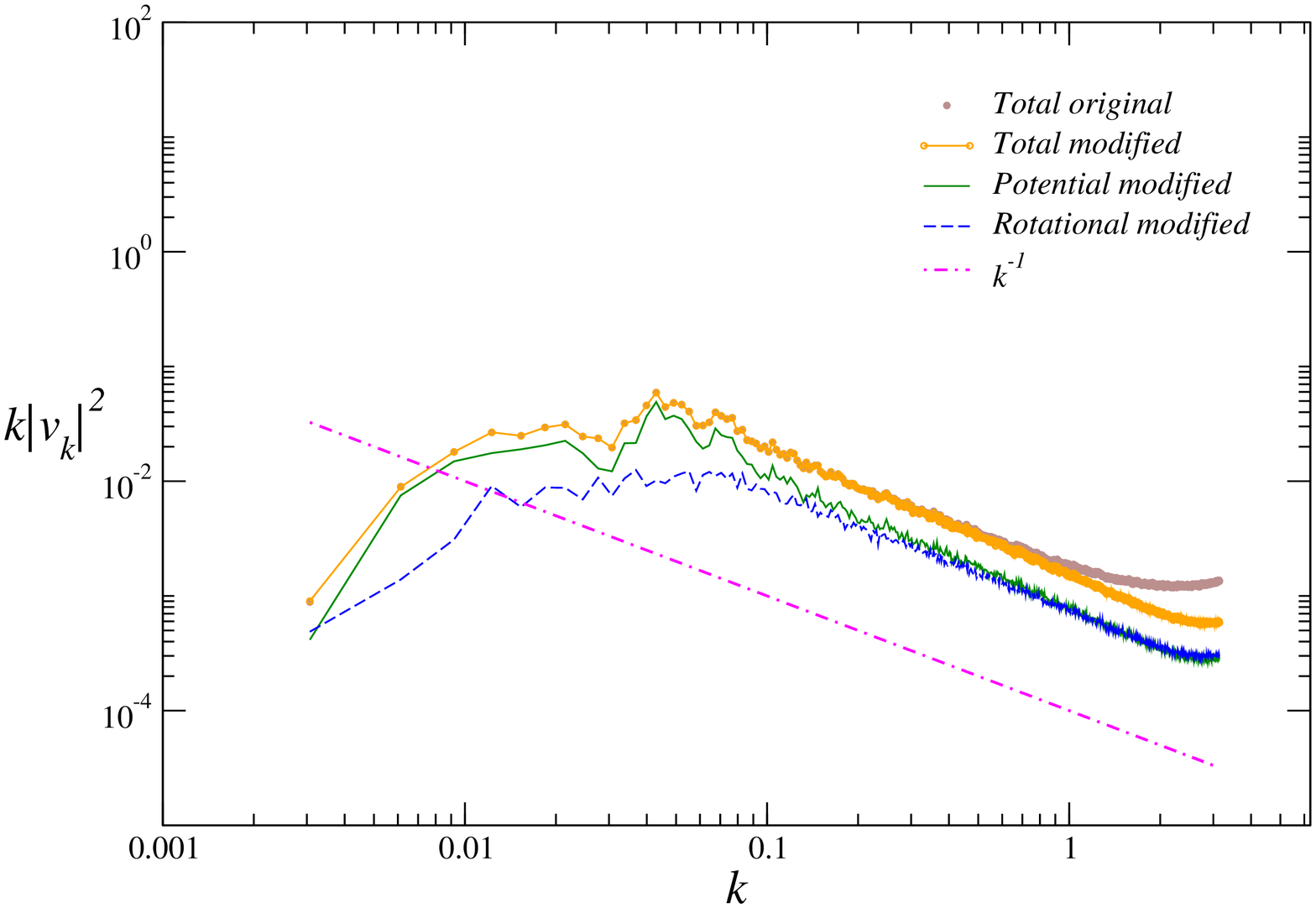}
  \includegraphics[width=8.cm]{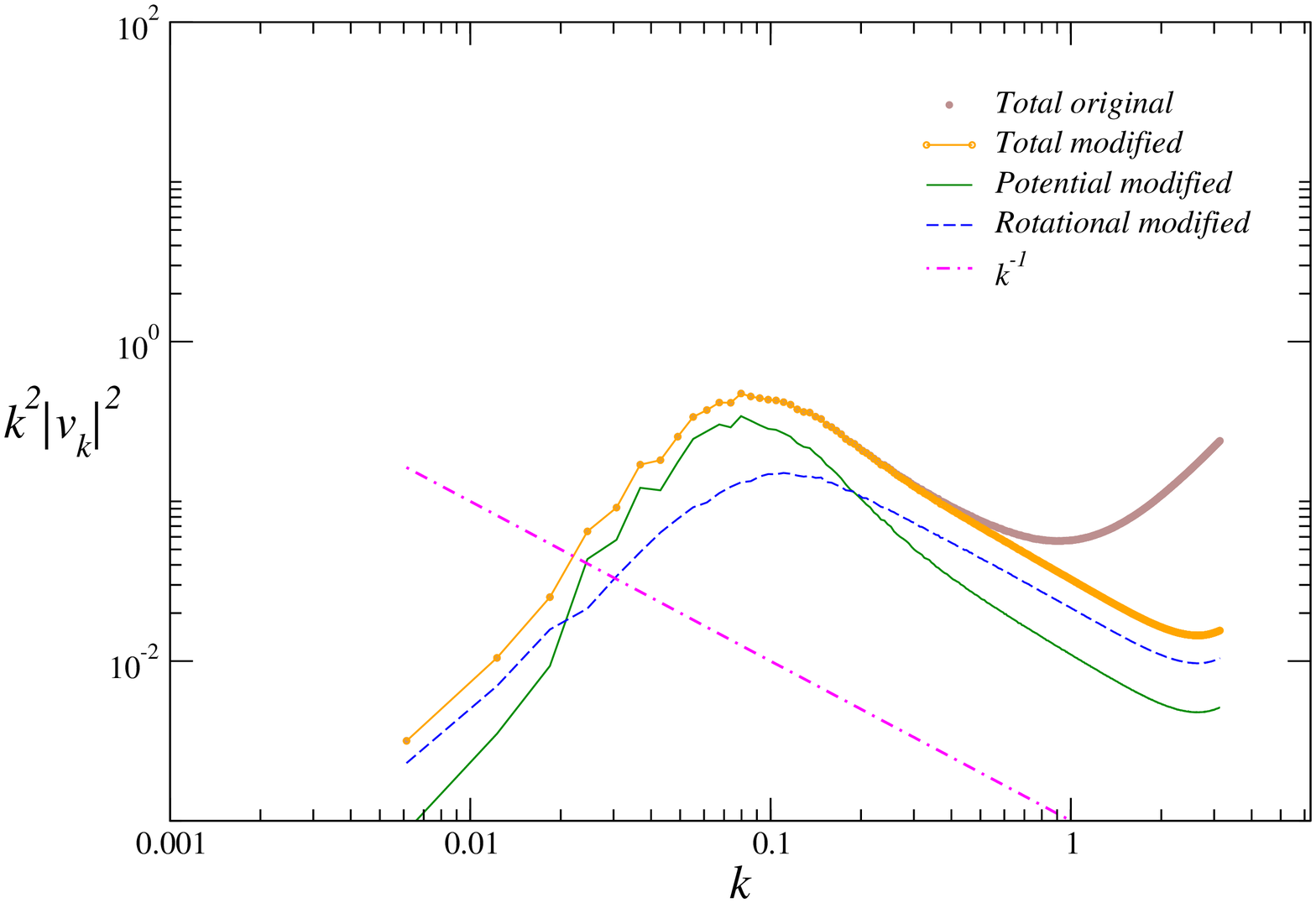}\\
  \caption{Steady-state power spectra of 2D (upper panel) and 3D (lower panel)
turbulence. The total flow, potential flow and rotational flow spectra are
plotted separately. The original spectra have unphysical upturns at the
highest wavenumber due to the grid-size singular flow velocity of vortex
that cannot be faithfully captured by the grids. Upon removal of these
unphysical pixels, the spectrum shows the almost perfect $k^{-1}$ power law
in both 2D and 3D turbulence.
They also exhibit energy equi-partition on small scales.
}
  \label{fig:2D_Equi_Partition}
 \end{center}
\end{figure}

Return to the turbulence spectrum. We note that vortices are line
objects when viewed in 3D. The fluid velocity is dominated by
local 2D flows around the vortex lines. Therefore when the vortex
lines are sparsely distributed in space, the vortex flow velocity
has strong short-distance correlation transverse
to the line but long-distance correlation along the line, leading
to similar 2D and 3D velocity spectra in high $k$.  In fact this $k^{-1}$
spectrum may be derived from the spectrum of a single vortex. The transverse
velocity around a vortex $|{\bf v}_\perp|$ scales as $|{\bf
r}|^{-1}$. Hence its Fourier component scales as $k^{-1}$, and the
spectrum as $k^{-1}$. In reality, there are a number of point
vortices and vortex loops in 2D and 3D. (See Figs.(3) and (4) for
real-space 2D and 3D images of vortices.) If these vortices are
weakly correlated, the total spectrum is simply the sum of all
individual vortex spectra and hence will still obey the $k^{-1}$ power
law. Indeed, since the 2D point vortex occupies a 2D area of
measure zero, so does the 3D loop vortex a zero volume, the
vortex-vortex two-point correlation function is therefore
overwhelmingly dominated by the zero-separation contribution, and
the resulting spectrum remains to be $k^{-1}$.

Nevertheless, the finite-separation contribution to the vortex-vortex
two-point correlation should still exist in this system.  It is simply due to the
strong zero-separation correlation that the finite-separation
correlation becomes not obvious.  For
example, the 2D vortex pair should have kept some memory of its
counterpart after pair separation, and therefore they should be
correlated. In 3D the vortex-loop element should also be
correlated with other elements in the same loop. Correlation of
different loops, once exists, should be an indication of non-trivial
dynamics in action.

To uncover this weak finite-separation correlation,
we treat every point vortex in 2D as either
a signed point or an unsigned point and examine the two-point
correlation function of these points. It is found that for the
unsigned case, the vortex correlation $\xi(\Delta r)$ does not
show any power-law relation with $\Delta r$, where $\Delta r$ is
the separation of two points. For the signed case, the two-point
correlation $\eta(\Delta r)$ does however exhibit the screening
effect, where at small separation the correlation is negative and
finite but the correlation abruptly drops to zero at some critical
separation. In fact the signed vortex correlation can be fit by a
Gaussian. The critical separation, or the variance of the
Gaussian, represents the mean vortex separation, which is
determined by the initial flow energy with larger flow energy
exciting more vortices and therefore yielding smaller vortex mean
separation. Figure (6) shows such 2D two-point correlation of signed and
unsigned cases.

For 3D finite-separation correlation , we also consider the vortex line as
either a directional line or a non-directional line. The directional line correlation
is defined as $\eta(\Delta r)\equiv \langle d{\bf l}_1({\bf
r})\cdot d{\bf l}_2({\bf r}+{\bf \Delta r})\rangle$, and the
nondirectional line correlation as $\xi(\Delta r)\equiv\langle
dl_1({\bf r}) dl_2({\bf r}+{\bf\Delta r})\rangle$. It turns out that
the nondirectional line-line correlation $\xi$ obeys a universal
$\Delta r^{-2}$ power law.  Although the directional
line-line correlation $\eta$ also obeys $\Delta r^{-2}$ power law
at small separation, it abruptly
drops to zero at a finite separation. The directional
line-line correlation $\eta$ reveals the average size of the
vortex loop at the separation when the correlation drops to zero.
Figure (7) depicts the 3D line-line correlation.
Comparing these correlation functions in 2D and 3D cases, we find that the dimensionality
plays a role in the final states. Two-dimensional quantum turbulence
in our model appears farther from a relaxed state than the 3D case since small-scale mixing
in 3D turbulence is well in place in the final steady state that yields a power-law correlation.

\begin{figure}
 \begin{center}
  \leavevmode
  \vspace{1cm}
  \includegraphics[width=8cm,angle=0]{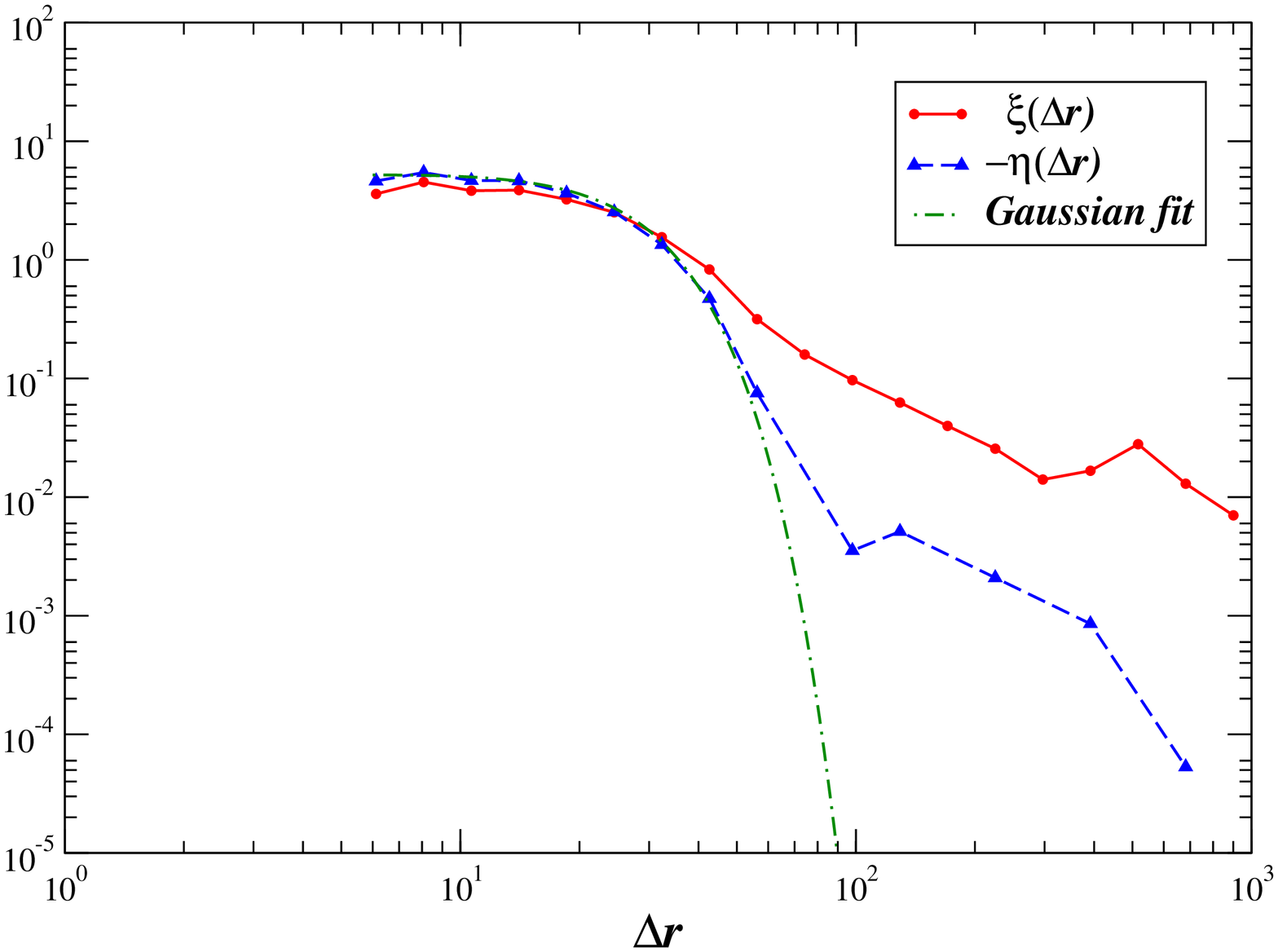}
  \caption{Two-point correlation functions of signed point vortices
$\eta$ and unsigned point vortices $\xi$. The signed vortex correlation
$\eta$ can be fit by a Gaussian and the variance of the Gaussian represents the
vortex mean separation.}\label{fig:2_Point_correlation}
 \end{center}
\end{figure}

\begin{figure}
 \begin{center}
  \leavevmode

  \includegraphics[width=8cm,angle=0]{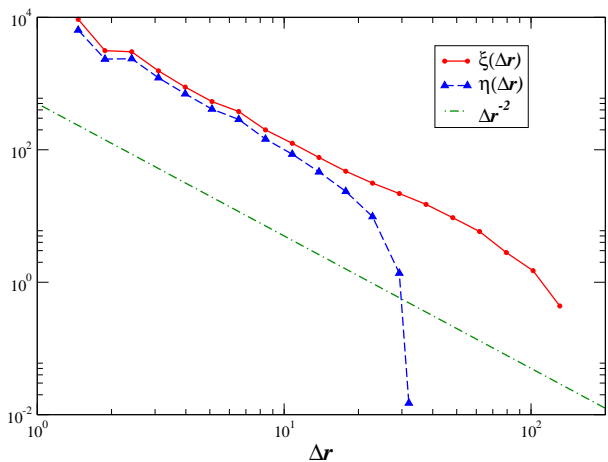}\\
  \caption{Line-Line correlation of directional vortex lines
$\eta$ and non-directional cortex lines $\xi$. Both follow
a power law $\Delta r^{-2}$ at small separation.
}\label{fig:2_Point_correlation_3D}
 \end{center}
\end{figure}

We now turn to the upturn at the highest-$k$ in the energy
spectrum. It is actually caused by numerical errors in the finest
grid at the vortex. The velocity field is singular at the vortex,
whose exact location can be anywhere within the finest grid,
giving unequal weights to the grids immediately surrounding the
vortex. When the vortex is much closer to a particular grid, the
velocity at that grid will be much larger than other nearby grids.
Therefore, this grid will appear as an erroneous Dirac
$\delta$-function, thereby giving an upturn with the first power
of $k$ in the spectrum at the smallest scale. We suppress this
problem by limiting the magnitude of velocity to an appropriate
upper bound so as to eliminate the contribution of the said
$\delta$-functions. (Only 20-30 grids are processed
among a total of $2048^2$ grids that contains about 100 point vortices in the 2D case,
and the 3D case also has a similar ratio of processed to unprocessed grids.)
Figure (5) also shows the spectra
after this regularization procedure. We emphasize that such a grid-scale
error is unavoidable since the local physics is singular; such an
error can accumulate over time when evolving a nonlinear schroedinger
equation numerically. Only in this
soluble model adopted here can the error in the solution be
controllable and does not accumulate.

Quantum vortices are long-lived coherent structures in
dissipationless turbulence. Once created, they can only come to
destruction through pair annihilation in 2D and through
shrinkage of ring vortex in 3D. When vortex lines are sparsely
distributed in space, vortex collisions are rare events, thereby
making them long-lived objects. Vortex-line reconnection arises
in 3D from vortex collision and is a more common
phenomenon than vortex annihilation.  Details of 3D vortex-line
reconnection are depicted in Fig.(8).
Also clear in Fig.(8) is the existence of helical interference fringes
along the cores of 3D vortex lines.  These fringes appear to vary rather slowly and
are the $m$(mode number around the tube)$=2$ density fluctuations of varying
$n$(mode number along the tube).  We mentioned previously that the density profile
cut across a vortex-tube is generally elliptical near the null, and this is consistent with
the $m=2$ density fluctuation observed at the 3D vortex core.  We also note that
Kelvin waves are the $m=1$ fluctuations and oscillate due to the "tension" of a vortex
line.  Thus, Kelvin waves appear qualitatively different from these $m=2$ fringes.

\begin{figure}
 \begin{center}
  \leavevmode
  \vspace{0.5cm}
  \includegraphics[width=6cm]{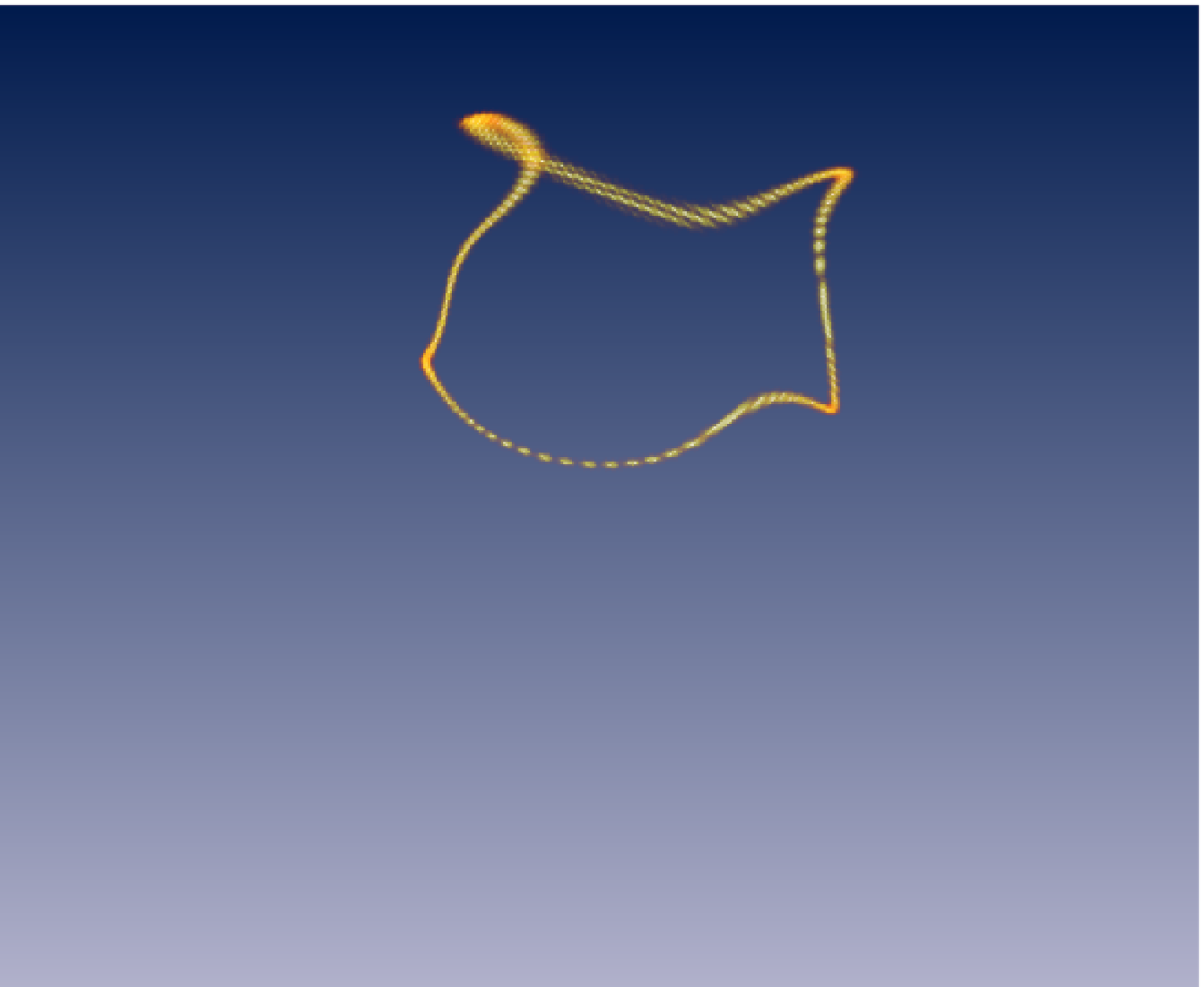}
  \vspace{0.5cm}
  \includegraphics[width=6cm]{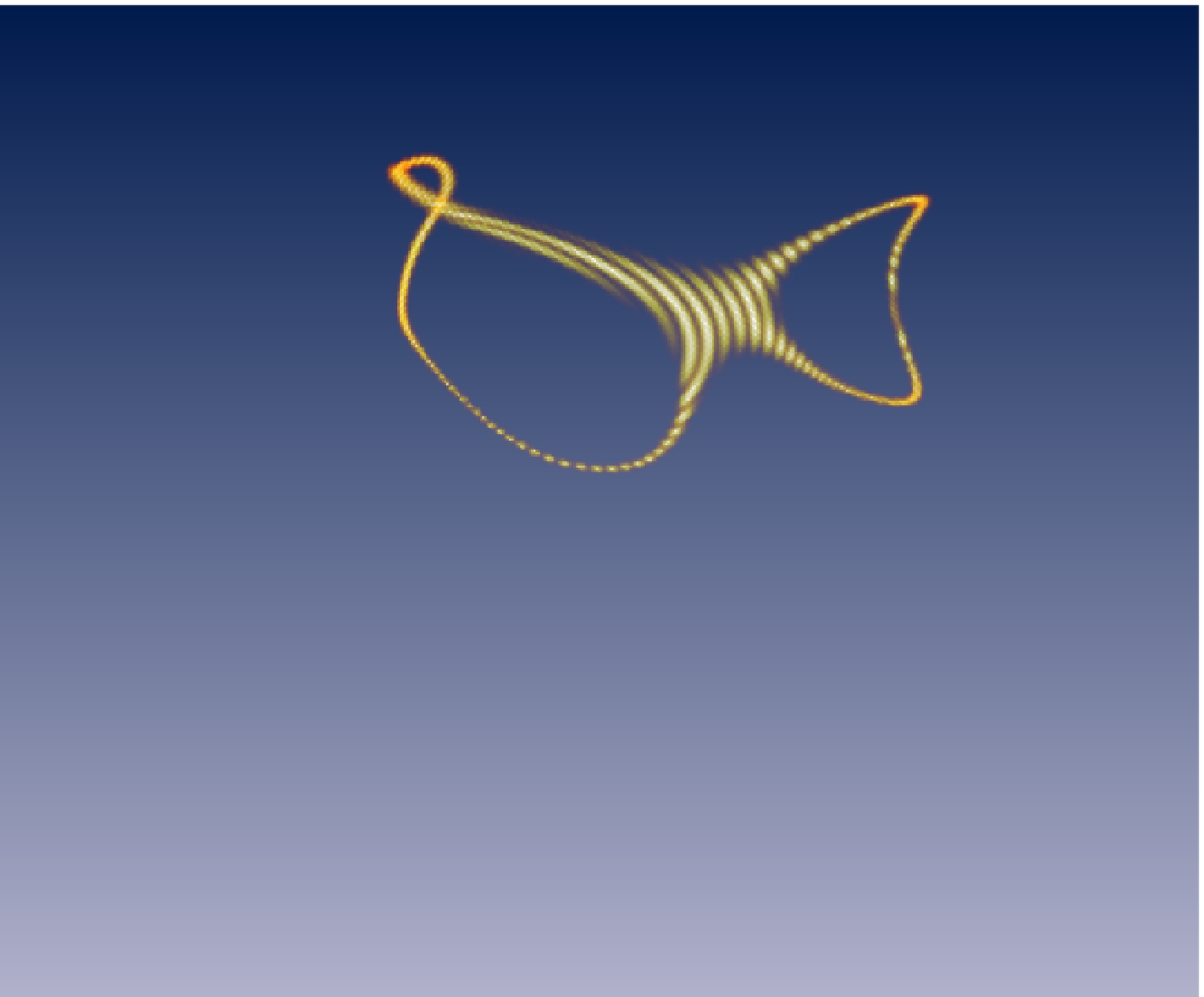}
  \vspace{0.5cm}
  \includegraphics[width=6cm]{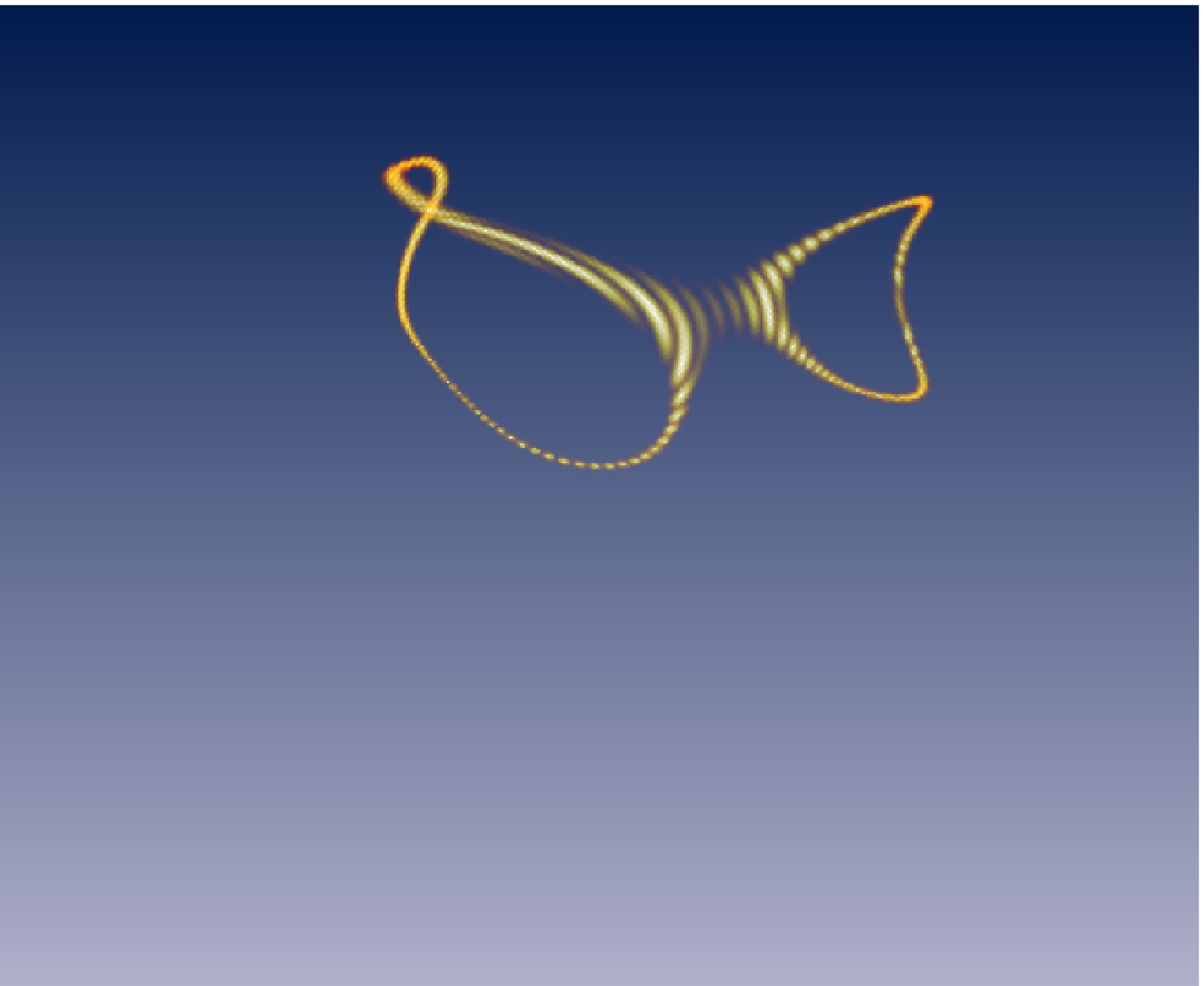}
  \vspace{0.5cm}
  \includegraphics[width=6cm]{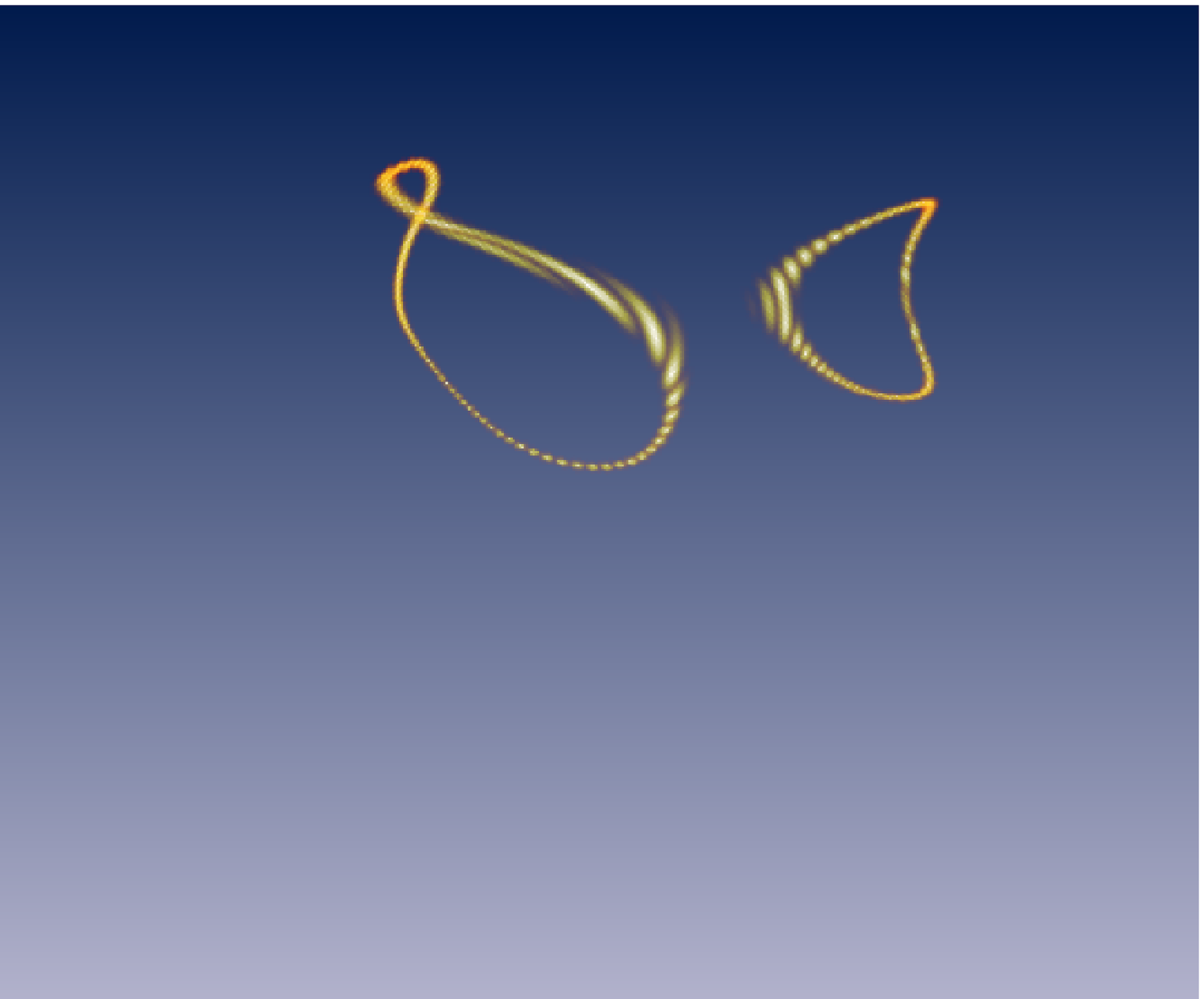}\\
  \caption{3D reconnection of vortex loop, which is the major activity for the
maintenance of steady-state quantum turbulence. The evolution
proceeds from the upper to bottom panels. We also note the
small-scale helical fringes in existence along the cores
of vortex lines. These helical fringes appear at the very core of a vortex tube,
as they can only be captured when the constant-density surfaces
assume an extremely low value, lower than that used to plot
Fig.4.}
  \label{fig:3D_reconnection}
 \end{center}
\end{figure}

In 2D quantum turbulence, inverse cascade has not been observed in
our system. We explore this issue by having the energy of initial
potential flow concentrating on the mid scale, hoping to provide
an ample space on the large scale for inverse cascades to take
place. However, the large-scale flow energy in fact does not
change throughout the evolution, indicating the non-existence of inverse
cascade in this model.  Treating unsigned vortices as
individual points, we again examine the vortex-vortex correlation,
aiming to find out whether these points have any tendency to
cluster when these point vortices are created on an intermediate
scale. Like the situation explored previously, we find that 2D
vortices do not develop long-range correlation. This aspect is in
great contrast to classical 2D fluid turbulence, where vortices of
the same sign tend to aggregate to form large vortex clusters.

\section{Conclusions}

By exploring the exactly soluble Schroedinger equation for
free-particles, we have derived the following generic results in
this work.  These results can also, or has been observed to, arise
from a nonlinear Schroedinger system. (1) Quantum vortex has a
axisymmetric rotational velocity but a non-axisymmetric potential
velocity, both of which diverge near the vortex as $r^{-1}$. (2)
Whenever vortices appear in the system, the density and momentum
equations of the fluid representation will become inadequate to
evolve the system, and it requires a third equation, the vortex
equation of motion, for the full description of evolution. (3)
Unlike classical vortex, quantum vortex is a wave that is not
frozen into the matter and it propagates at a velocity different from
the local matter velocity. (4) The motion of quantum vortex is not
entirely induced by other vortices and hence its propagation
velocity is different from the rotational component of local matter
velocity. (5) Equip-partition between rotational flow energy and
potential flow energy can be established on small scales in fully
developed quantum turbulence.

Specific results for our 2D and 3D free-particle models are that
(a) starting with an initial potential flow, turbulence cascades
toward small-scale is observed prior to rotational flows are
excited, and near the transition to the excitation of rotational flow the
energy spectra of potential
flow turbulence in 2D and 3D obey different power laws; (b) the energy
spectra of fully developed 2D and 3D turbulence with vortex excitation
both obey the same
$k^{-1}$ power law, which corresponds to the power spectrum of a single vortex;
(c) $\Delta r^{-2}$ power-law vortex-vortex correlation is observed only in 3D
but not in 2D turbulence; (d) there is no evidence of inverse
cascade in 2D turbulence; (e) there is no evidence of Kolmogorov energy
cascade in either 2D or 3D turbulence.

By comparing our general and specific results, it is evident that
small-scale features derived from this linear Schroedinger equation can be in
common with those derived from a nonlinear Schroedinger equation. But the
large-scale features, such as vortex-vortex correlation, cannot, because the
wave nonlinearity and dissipation, which are lacking
in this linear model, must
play an essential role in shaping the global dynamics. Along this line,
we note that the steady-state turbulence spectrum is maintained not by the
conventional cascade processes in this model, since no dissipation is involved.
In fact, energy cascade is only observed in the initial potential flow phase as a
transient before vortices are excited.
Once vortices are fully excited, the steady-state $k^{-1}$ energy spectrum
found in this model only reveals the spectrum of a collection of individual vortices
and contains no information about correlation among different vortices.

Whether it is the dissipation or the wave nonlinearity that is
the prime factor for producing the long-range vortex-vortex
correlation is an important issue for understanding the actual quantum
turbulence. In fact, in the presence of small-scale
dissipation the $k^{-1}$ steady-state spectrum must fail, as it would have
yielded a diverging dissipation rate.
Whether the Kolmogorov spectrum may be established in a
dissipative linear Schroedinger system remains
to be investigated.
In the case of a nonlinear Schroedinger system, Kelvin wave cascades are found
to play a central role in coupling to dissipation [8,9]. 
On this particular issue, our observation for a vortex loop of
a linear Schroedinger system is that it can
undergo large-amplitude loop distortion so that the loop is self pinched
and reconnects to form two smaller loops (c.f., Fig. (8)).  This reconnection process
can couple to dissipation efficiently if some small scale dissipation is included in the system.   
On the other hand, despite Kevin waves also ride on a vortex loop and
can give rise to large-amplitude loop distortion, the picture of Kelvin wave cascade
from long waves to short waves and finally to dissipation on a vortex loop is quite different 
from our observation of vortex loop dynamics of a linear Schroedinger system.

Finally, we should mention that the present dissipationless linear
system in a periodic box can be recurrent in a finite time when the longest wave in
the system oscillates
over one period.  The steady state turbulence presented in this work
is found to be already established within a small fraction (10$\%$) of the recurrent time.
The Poincare recurrence time of this model is proportional
to the square of the periodic box size, $L^2$.
Despite the dynamics of this system may appear chaotic, it is not quite a chaotic system
due to the algebraic, instead of an exponential,
dependence of the recurrence time on the degrees of freedom.

\vspace{2cm}
\begin{acknowledgments}
This work is supported in part by the grant, NSC97-2628-M002-008-MY3,
from National Science Council of Taiwan.
\end{acknowledgments}

\vspace{2cm}
\bibliography{Reference}
\noindent[1] R. P. Feynman, Progress in Low Temperature Physics
(North-Holland, Amsterdam, 1955), Vol. I.

\noindent[2] J. Yepez, G.Vahala, L. Vahala,3 and M. Soe, Phys. Rev. Lett. 103, 084501 (2009);
C. Nore, M. Abid, and M. E. Brachet, Phys. Rev. Lett. 78,
3896 (1997);
C. F. Barenghi, Physica (Amsterdam) 237D, 2195 (2008);
E. Kozik and B. Svistunov, Phys. Rev. Lett. 92, 035301
(2004); V. S. L'Vov, S.V. Nazarenko, and O. Rudenko,
Phys. Rev. B 76, 024520 (2007); W. F. Vinen,
M. Tsubota, and A. Mitani, Phys. Rev. Lett. 91, 135301
(2003); E. Kozik and B. Svistunov, Phys. Rev. B 77, 060502(R) (2008)

\noindent[3] T. Chiueh, Phys. Rev. E 57, 4150 (1998)

\noindent[4] A.A. Abrikosov, Rev. Mod. Phys. 76, 975 (2004)

\noindent[5] M.R. Matthews, et al., Phys. Rev. Lett. 83, 2498 (1999)

\noindent[6] C.N. Weiler, et al., Nature, 455, 948 (2008)

\noindent[7] D.V. Freilich, et al., Science, 329, 1182 (2010)

\noindent[8] W.F. Vinen, J. Low Temp. Phys. 145, 7 (2006)

\noindent[9] M. Kobayashi and M. Tsubota, Phys. Rev. Lett. 94,
065302 (2005); M. Kobayashi and M. Tsubota, J. Phys. Soc. Jpn. 74, 3248 (2005);
C. F. Barenghi, Physica (Amsterdam) 237D, 2195 (2008);
T. P. Simula, T. Mizushima, and K. Machida, Phys. Rev. Lett. 101, 020402 (2008);
A.L. Fetter, 2004 Phys. Rev. A, 69, 043617 (2004); V. Bretin, P. Rosenbusch,
F. Chevy, G.V Shlyapnikov, and J. Dalibard, Phys. Rev. Lett., 90,100403 (2003)

\noindent[10] I. Bialynicki-Birula, Z. Bialynicka-Birula, and C. Sliwa1, Phys.Rev. A, 61, 032110 (2000)

\noindent[11] T.P. Woo and T. Chiueh, ApJ, 697, 850 (2009)

\noindent[12] M. A. Green, C. W. Rowley, and G. Haller, J Fluid Mech., 572, 111 (2007)

\noindent[13] Z.S. She and E. Leveque, Phys. Rev. Lett. 72, 336 (1994); Z.S. She and
E.C. Waymire, Phys. Rev. Lett, 74, 262 (1995); D. Berengere, Phys. Rev. Lett. 73, 959 (1994)

\noindent[14] R. J. Donnelly, Quantized Vortices in Helium II
(Cambridge University Press, Cambridge, England, 1991)

\end{document}